\documentclass[12pt]{article}

\newif\ifnew
\newtrue

\newcommand{\new}[1]{\ifnew #1\fi}

\begin{document}

\title{       \new{Using Information Theory Approach to  Randomness Testing}
 \footnote {
\new{ The authors were  supported by INTAS
     grant no. 00-738 and Russian Foundation for Basic Research under Grant no. 03-01-00495.}
 } }
\author{ B. Ya. Ryabko and V.A. Monarev
}
\date{}
\maketitle

\begin{abstract}
We address the problem of detecting deviations of binary sequence
from randomness,which is very important for random number (RNG)
and pseudorandom number generators (PRNG). Namely, we consider a
null hypothesis $H_0$ that a given bit sequence is generated by
Bernoulli source with equal probabilities of 0 and 1 and the
alternative hypothesis $H_1$ that the sequence is generated by a
stationary and ergodic source which differs from the source under
$H_0$. We show that data compression methods can be used as a
basis for such testing and describe two new tests for randomness,
which are based on ideas of universal coding. Known statistical
tests and suggested ones are applied for testing  PRNGs. Those
experiments show that the power of the new tests is greater than
of many known algorithms.

\end{abstract}

\newpage

\section{Introduction }

The randomness testing of random number and pseudorandom number
generators is used for many purposes including cryptographic,
modeling and simulation applications; see, for example, Knuth,
1981; L'Ecuyer, 1994; Maurer,1992; Menezes A. and others, 1996.
For such applications a required bit sequence should be true
random, i.e., by definition, such a sequence could be interpreted
as the result of the flips of a "fair" coin with sides that are
labeled "0" and "1" (for short, it is called a random sequence;
see Rukhin and others, 2001). More formally, we will consider the
main hypothesis $H_0$ that a bit sequence is generated by the
Bernoulli source with equal probabilities of 0's and 1's.
Associated with this null hypothesis is the alternative
hypothesis $H_1$ that the sequence is generated by a stationary
and ergodic source which generates letters from $\{0,1\}$ and
differs from the source under $H_0$.

In this paper we will consider some tests which are based on
results and ideas of Information Theory and, in particular, the
source coding theory. First, we  show that a universal code can be
used for randomness testing. (Let us recall that, by definition,
the universal code can compress a sequence asymptotically till the
Shannon entropy per letter when the sequence is generated by a
stationary and ergodic source). If we take into account that the
Shannon per-bit entropy is maximal (1 bit) if $H_0$ is true and is
less than 1 if $H_1$ is true (Billingsley, 1965; Gallager, 1968),
we see that it is natural to use this property and universal codes
for randomness testing because, in principle,  such a test can
distinguish each deviation from randomness, which can be described
in a framework of the stationary and ergodic source model. Loosely
speaking, the test rejects $H_0$ if a binary sequence can be
compressed by a considered universal code (or a data compression
method.)

It should be noted that the idea to use the compressibility as a
measure of randomness has a long history in mathematics. The point
is that, on the one hand, the problem of randomness testing is
quite important for practice, but, on the other hand, this problem
is closely connected with such deep theoretical issues as the
definition of randomness, the logical  basis of probability
theory, randomness and complexity, etc; see Kolmogorov, 1965; Li
and Vitanyi, 1997; Knuth, 1981; Maurer,1992. Thus, Kolmogorov
suggested to define the randomness of a sequence, informally, as
the length of the shortest program, which can create the sequence
(if one of the universal Turing machines is used as a computer).
So, loosely speaking, the randomness (or Kolmogorov complexity) of
the finite sequence is equal to its shortest description. It is
known that the Kolmogorov complexity is not computable and,
therefore, cannot be used for randomness testing. On the other
hand, each lossless data compression code can be considered as a
method for upper bounding the Kolmogorov complexity. Indeed, if
$x$ is a binary word, $\phi$ is a data compression code  and
$\phi(x)$ is the codeword of $x$, then the length of the codeword
$|\phi(x)|$ is the upper bound for the Kolmogorov complexity of
the word $x$. So, again we see that the codeword length of the
lossless data compression method can be used for randomness
testing.

In this paper we suggest tests for randomness, which are based on
results and ideas of the source coding theory.

Firstly, we show how to build a test basing on any data
compression method and give some examples of application of such
test to PRNG's testing. It should be noted that data compression
methods were considered as a basis for randomness testing in
literature. For example, Maurer's Universal Statistical Test,
Lempel-Ziv Compression Test and Approximate Entropy Test are
connected with universal codes and are quite popular in practice,
see, for example, Rukhin  and others, 2001. In contrast to known
methods, the suggested approach gives a possibility to make a test
for randomness, basing on any lossless data compression method
even if a distribution law of the codeword lengths is not known.

Secondly, we describe two new tests, conceptually connected with
universal codes. When both tests are applied, a tested sequence
$x_1 x_2 ... x_n$ is divided into subwords  $x_1 x_2 ... x_s,$
$\:x_{s+1} x_{s+2} ... x_{2s},\: \ldots ,\,$ $s\geq 1,$ and the
hypothesis $H^*_0$ that the subwords obey the uniform distribution
(i.e. each subword is generated with the probability $2^{-s}$) is
tested against $H^*_1 =\neg H^*_0$. The key idea of the new tests
is as follows. All subwords from the set $ \{0,1\}^s $ are ordered
and this order changes after processing each subword $\:x_{j s+1}
x_{j s+2} ... x_{(j+1)s}, \, j= 0,1, \ldots $ in such a way that,
loosely speaking,  the more frequent subwords have small ordinals.
When the new tests are applied, the frequency of different
ordinals are estimated (instead of frequencies of the subwords as
for, say, chi- square test).

The natural question is how to choose the block length $s$ in such
schemes. We show that, informally speaking, the block length $s$
should be taken quite large  due to the existence of  so called
{\it two-faced processes}. More precisely, it is shown that for
each integer $s^*$ there exists such a process $\xi$ that for each
binary word $u$ the process $\xi$ creates $u$ with the probability
$2^{-|u|}$ if the length of the $u$  ($|u|$) is less than or equal
to  $s^*$, but, on the other hand,  the probability distribution
$\xi(v)$ is very far from uniform if the length of the words $v$
is greater than $s^*.$ (So, if we use a test with the block length
$s \leq s^*,$ the sequences generated by $\xi$ will look like
random, in spite of $\xi$ is far from being random.)

The outline of the paper is as follows. In Section 2 the general
method for construction randomness testing algorithms basing on
lossless data compressors is described. Two new tests for
randomness, which are based on constructions of universal coding,
as well as the two-faced processes, are described in the Section
3. In Section 4 the new tests are experimentally  compared with
methods from  " A statistical test suite for random and
pseudorandom number generators for cryptographic applications",
which was recently suggested by Rukhin  and others, 2001. It turns
out that the new tests are more powerful than known ones.
\section{Data compression methods as a basis for randomness testing }

\textbf{2.1. Randomness testing based on data compression}

Let $A$ be a finite alphabet and $A^n$ be the set of all  words of
the length $n$ over $A$, where $n$ is an integer. By definition,
$A^* =\bigcup_{n=1}^\infty A^n $ and $A^\infty$ is the set of all
infinite words $x_1x_2 \ldots $ over the alphabet $A$. A data
compression method (or code) $\varphi$ is defined as a set of
mappings $\varphi_n $ such that $\varphi_n : A^n \rightarrow \{
0,1 \}^*,\, n= 1,2, \ldots\, $ and for each pair of different
words $x,y \in A^n \:$ $\varphi_n(x) \neq \varphi_n(y) .$
Informally, it means that the code $\varphi$ can be applied for
compression of each message of any length $n, n
> 0 $ over alphabet $A$ and the message can be decoded if
its code is known.

Now we can describe a statistical test which can be constructed
basing on any code $\varphi$.  Let $n$ be an integer and
$\hat{H}_0$ be a hypothesis that the words from the set $ A^n $
obey the uniform distribution, i.e., $p(u)= |A|^{-n}\, $ for each
$ \, u \in \{0,1\}^n .$ (Here and below $|x|$ is the length if $x$
is a word, and the number of elements if $x$ is a set.)  Let a
required  level of significance (or a Type I error) be $\alpha ,\,
\alpha \in (0,1).$ The following  main idea of a suggested test is
quite natural: The well compressed words should be considered as
non- random and $\hat{H}_0$ should be rejected. More exactly, we
define a critical value of the suggested test by
\begin{equation}\label{cr}
t_\alpha = n \log |A| - \log (1/ \alpha) - 1\,.
\end{equation}
(Here and below $\log x = \log_2 x$.)

Let $u$ be a word from $A^n$. By definition, the hypothesis
$\hat{H}_0$  is accepted if $ |\varphi_n (u) | > t_\alpha $ and
rejected, if $ |\varphi_n (u) | \leq t_\alpha .$  We denote this
test by $\Gamma_{\alpha,\,\varphi}^{(n)}.$

\textbf{Theorem 1.} { \it For each integer $n$ and  a code
$\varphi$, the Type I error of the described test
$\Gamma_{\alpha,\,\varphi}^{(n)}$ is not larger than $\alpha .$ }

\emph{Proof} is given in Appendix.

\textbf{ Comment 1}. The described test can be modified in such a
way that the Type I error will be equal to $\alpha.$ For this
purpose we define the set $A_\gamma$ by $$ A_\gamma = \{x: x \in
A^n\: \: \& \:\;|\varphi_n(x)| = \gamma \} $$ and an integer $g$
for which the two following inequalities are valid:
\begin{equation}\label{s}  \sum_{j=0}^g
|A_j|\: \leq\,  \alpha |A|^n\, <\, \sum_{j=0}^{g+1} |A_j| \,.
\end{equation} Now the modified test can be described as
follows:

If for $x \in A^n\;\; |\varphi_n(x)| \leq g\:\; $ then $\hat{H}_0$
is rejected, if $|\varphi_n(x)| >  (g+1) \:$ then $\hat{H}_0$ is
accepted and if $|\varphi_n(x)| =  (g+1) \:$ the hypothesis
$\hat{H}_0$ is accepted with the probability $$ (\sum_{j=1}^{g+1}
|A_j|\, - \, \alpha |A|^n\,) / |A_{g+1}| $$ and rejected with the
probability $$ 1\:-\, (\sum_{j=1}^{g+1} |A_j|\, - \, \alpha
|A|^n\,) / |A_{g+1}| \,.$$ (Here we used a randomized criterion,
see for definition, for example, Kendall and  Stuart, 1961, part
22.11.) We denote this test by
$\Upsilon_{\alpha,\,\varphi}^{(n)}.$

\textbf{ Claim 1}. { \it For each integer $n$ and  a code
$\varphi$, the Type I error of the described test
 $\Upsilon_{\alpha,\,\varphi}^{(n)}$ is equal to $\alpha .$ }

\emph{Proof} is given in Appendix.

We can see that this criterion has the level of significance (or
Type I error) exactly $\alpha,$ whereas the  first criterion,
which is  based on critical value (\ref{cr}), has the level of
significance that could be less than $\alpha .$ In spite of this
drawback, the first criterion may be more useful due to its
simplicity. Moreover, such an approach gives a possibility to use
a data compression method $\psi$ for testing even in case where
the distribution of the length $|\psi_n(x)|, x \in A^n$ is not
known.

\textbf{ Comment 2.} We have considered  codes, for which
different words of the same length have different codewords (In
Information Theory sometimes such codes are called non- singular.)
Quite often a stronger restriction is required in Information
Theory. Namely, it is required that each sequence
$\varphi_n(x_1)\varphi_n(x_2) ...\varphi(x_r), r \geq 1,$ of
encoded words from the set $A^n, n\geq 1,$ can be uniquely decoded
into $x_1x_2 ...x_r$. Such codes are called uniquely decodable.
For example, let $A=\{a,b\}$, the code $\psi_1(a) = 0, \psi_1(b) =
00, $  obviously, is non- singular, but is not uniquely decodable.
(Indeed, the word $000$ can be decoded in both $ab$ and $ba.$) It
is well known in Information Theory that a code $\varphi$ can be
uniquely decoded if the following Kraft inequality is valid:
\begin{equation}\label{KRAFT}
\Sigma_{u \in A^n}\: 2^{- |\varphi_n (u) |} \leq 1\:,
\end{equation}
see, for ex., Gallager, 1968.

If it is known that the code is uniquely decodable, the suggested
 critical value (\ref{cr}) can be changed. Let us define \begin{equation}\label{cr2}
\hat{t}_\alpha = n \log |A| - \log (1/ \alpha) \,.
\end{equation}

Let, as before,  $u$ be a word from $A^n$. By definition, the
hypothesis $\hat{H}_0$  is accepted if $ |\varphi_n (u) | >
\hat{t}_\alpha $ and rejected, if $ |\varphi_n (u) | \leq
\hat{t}_\alpha .$ We denote this test by
$\hat{\Gamma}_{\alpha,\varphi}^{(n)}.$

\textbf{Claim 2.} { \it For each integer $n$ and  a uniquely
decodable code $\varphi$, the Type I error of the described test
$\hat{\Gamma}_{\alpha,\varphi}^{(n)}$ is not larger than
$\alpha.$}

\emph{Proof} is given in Appendix.

So, we can see from (\ref{cr}) and (\ref{cr2}) that the critical
value is larger, if the code is uniquely decodable. On the other
hand, the difference is quite small and (\ref{cr}) can be used
without a large loose of the test power even in a case of the
uniquely decodable codes.

It should not be a surprise that the level of significance (or a
Type I error) does not depend on the alternative hypothesis $H_1,$
but, of course,  the power of a test (and the Type II error) will
be  determined by $H_1.$

The examples of testing by real data compression methods will be
given in Section 4.

\textbf{ 2.2. Randomness testing based on universal codes. }

  We will consider  the  main
hypothesis $H_0$ that the letters of a given sequence $x_1x_2
...x_t, \, x_i \in A,\, $ are independent and identically
distributed (i.i.d.) with equal probabilities of all $a \in A $
and the alternative hypothesis $H_1$ that the sequence is
generated by a stationary and ergodic source, which generates
letters from $A$ and differs from the source under $H_0$. (If $A=
\{0,1\}$, i.i.d. coincides with Bernoulli source.) The definition
of the stationary and ergodic source and the Shannon entropy of
such sources can be found in Billingsley, 1965, and Gallager,
1968.

We will consider statistical tests, which are based on universal
coding and universal prediction. First we define a universal code.

By definition,  $\varphi$ is a universal code if for each
stationary and ergodic source  (or a process) $\pi$ the following
equality is valid with probability 1 (according to the measure
$\pi \,) $

\begin{equation}\label{un}
 \lim_{n \rightarrow \infty}  (|\varphi_n(x_1 ... x_n)|) /
n = h(\pi)\,,
\end{equation}
where  $h(\pi)$ is the Shannon entropy. ( Such codes exist, see
Ryabko, 1984.) It is well known in Information Theory that
$h(\pi)= \log |A|$ if $H_0$ is true, and $h(\pi)< \log |A|$ if
$H_1$ is true, see, for ex., Billingsley, 1965; Gallager, 1968.
From this property and (\ref{un}) we can easily yield  the
following theorem.

\textbf{Theorem 2.} { \it Let $\varphi$ be a universal code,
$\alpha \in (0,1)$ be a level of significance and a sequence
$x_1x_2 ...x_n, \, n \geq 1, \, $ be generated by a stationary
ergodic source $\pi$. If the described above test
$\Gamma_{\alpha,\,\varphi}^{(n)}$ is applied for testing $H_0$
(against $H_1$), then, with probability 1, the Type I error is not
larger than $\alpha$, and the Type II error goes to 0, when
$n\rightarrow \infty$. }

So, we can see that each good universal code can be used as a
basis for randomness testing. But converse proposition is not
true. Let, for example, there be a code, whose codeword length is
asymptotically equal to $ (0.5+ h(\pi) / 2 ) $ for each source
$\pi$ (with probability 1, where, as before, $h(\pi)$ is the
Shannon entropy). This code is not good, because its codeword
length does not tend to the entropy, but, obviously, such code
could be used as a basis for a test of randomness. So, informally
speaking, the set of tests is larger than the set of universal
codes.

Note that the close problems were considered by Bailey (1974), who
obtained many important results in this field.

\section{Two new tests for randomness and two-faced processes }

Firstly, we suggest two tests which are based on ideas of
universal coding, but they are described in such a way that can be
understood without any knowledge of Information Theory.

\textbf{ 3.1. The "book stack" test }

Let, as before, there be given an alphabet $A= \{a_1, ... , a_S
\},$ a source, which generates letters from $A,$ and two following
hypotheses: the source is i.i.d. and $p(a_1)= ....= p(a_S) =
1/S\:$ ($H_0$) and $H_1 = \neg H_0.$ We should test the hypotheses
basing on a sample $x_1 x_2 \,... \,x_n,\, n\geq 1\,,\,$ generated
by the source. When the "book stack" test is applied, all letters
from $A$ are ordered from 1 to $S$ and this order is changed after
observing each letter $x_t$ according to the formula

\begin{equation}\label{nu}
\nu^{t+1}(a)=\cases{1,&if $x_t = a\,$;\cr
           \nu^t(a)+1,&if $\nu^t(a) < \nu^t(x_t)$;\cr
           \nu^t(a), &if $ \nu^t(a) > \nu^t(x_t)$\, ,}
\end{equation}
where $\nu^t$ is the order after observing $x_1 x_2 \,... \,x_t,\,
t = 1\,,, ...\,, n\,,$ $\nu^1$ is defined arbitrarily. (For ex.,
we can define $\nu^1 = \{a_1, ... , a_S \}.$) Let us explain
(\ref{nu}) informally. Suppose that the letters of $A$ make a
stack, like a stack of books and  $\nu^1(a)$ is a position of $a$
in the stack. Let the first letter $x_1$ of the word $x_1 x_2
\,... \,x_n$ be $a$. If it takes $i_1-$th position in the stack
($\nu^1(a)= i_1$), then  take $a$ out of the stack and put it on
the top. (It means that the order is changed according to
(\ref{nu}).) Repeat the procedure with the second letter $x_2$
and the stack obtained, etc.

It can help to  understand the main idea of the suggested method
if we take into account that, if $H_1$ is true, then frequent
letters from $A$ (as frequently used books) will have relatively
small numbers (will spend more time next to the top of the stack).
On the other hand, if $H_0$ is true, the probability to find each
letter $x_i$ at each position $j$ is equal to $1/S$.

Let us proceed with the description of the test. The set of all
indexes  $ \{1, \ldots, S \} $ is divided into $r, r \geq 2,  $
subsets $A_1 = \{ 1,2,\ldots, k_1 \}, $ $ A_2 = \{ k_1+1,\ldots,
k_2 \}, \ldots , A_r = \{ k_{r-1}+1,\ldots, k_r \}.$ Then, using
$x_1 x_2 \,... \,x_n$, we calculate how many $\nu^t(x_t),$ $
t=1,..., n,$ belong to a subset $A_k, k=1,..., r$. We define this
number as $n_k$ (or, more formally, $n_k = | \{ t : \nu^t(x_t) \in
A_k, t=1,\ldots, n \}|,  k=1,..., r .$) Obviously, if $H_0$ is
true, the probability of the event $ \nu^t(x_t) \in A_k$ is equal
to $ |A_j|/S.$ Then, using a "common" chi- square test we test the
hypothesis $\hat{H}_0= P\{  \nu^t(x_t) \in A_k \}= |A_k|/S $
basing on the empirical frequencies $n_1,\ldots,n_r$, against
$\hat{H}_1= \neg \hat{H}_0.$ Let us recall that the value
\begin{equation}\label{x2}
 x^2=\sum_{i=1}^{r}\frac{(n_i - n (|A_i|/S ) )^2}{n
(|A_i|/S )} \end{equation} is calculated, when  chi- square test
is applied, see, for ex., Kendall and  Stuart, 1961. It is known
that $x^2$ asymptotically follows the $\chi$-square distribution
with $(k-1)$ degrees of freedom ($\chi^2_{k-1}$) if $\hat{H}_0$ is
true. If the level of significance (or a Type I error)
 of the $\chi^2$
test is $\alpha, \alpha \in (0,1), $ the hypothesis $\hat{H}_0$ is
accepted when $x^2$ from (\ref{x2}) is less than the
\emph{$(1-\alpha)$ -value } of the $\chi^2_{k-1}$ distribution;
see, for ex.,  Kendall, Stuart, 1961.

We do not describe the exact rule  how to construct the subsets
$\{A_1, A_2, $ $  \ldots, $ $  A_r \}$, but we recommend to
perform some experiments for finding the parameters, which make
the sample size minimal (or, at least, acceptable). The point is
that there are many cryptographic and other applications where it
is possible to implement some experiments for optimizing the
parameter values and, then, to test hypothesis basing on
independent data. For example, in case of testing a PRNG it is
possible to seek suitable parameters using a part of generated
sequence and then to test the PRNG using a new part of the
sequence.

Let us consider a simple example. Let $A= \{a_1, \ldots , a_6 \},
$ $ r=2, A_1= \{a_1,a_2, a_3 \} , A_2=  \{a_4, a_5, a_6 \}, $ $
x_1 \ldots x_8 =$ $ a_3 a_6 a_3 a_3 a_6 a_1 a_6 a_1.$ If $\nu_1=
1, 2, 3, 4,$ $ 5,6 ,$ then $\nu_2= 3, 1, 2,  4, 5,6 ,$ $\nu_3= 6,
3, 1, 2, 4, 5 ,$ etc., and  $n_1 = 7, n_2 = 1.$ We can see that
the letters $ a_3 $ and $a_6$ are quite frequent and  the "book
stack" indicates this nonuniformity quite well. (Indeed, the
average values of $n_1$ and $n_2$ equal $4$, whereas the real
values are 7 and 1, correspondingly.)

Examples of practical applications of this test will be given in
Section 4, but here we make two notes. Firstly, we pay attention
to the complexity of this algorithm. The "naive" method of
transformation according to (\ref{nu}) could take the number of
operations proportional to $S,$ but there exist algorithms, which
can perform all operations in (\ref{nu}) using $O( \log S )$
operations. Such algorithms can be based on AVL- trees, see, for
ex., Aho,Hopcroft and Ulman, 1976.

The last comment concerns with the name of the method. The "book
stack" structure is quite popular in Information Theory and
Computer Science. In Information Theory this structure was firstly
suggested as a basis of an universal code by Ryabko, 1980, and
 was rediscovered by Bently, Sleator,  Tarjan, Wei in 1986, and
Elias in 1987 (see also a comment of Ryabko (1987) about a history
of this code). In English language literature this code is
frequently called as "Move-to-Front" (MTF) scheme  as it was
suggested by Bently, Sleator,  Tarjan and  Wei. Now this data
structure is used in a caching and many other algorithms in
Computer Science under the name "Move-to-Front". It is also worth
noting that the book stack was firstly considered by a soviet
mathematician M.L. Cetlin as an example of a self- adaptive
system in 1960's, see Rozanov, 1971.

\textbf{ 3.2. The order test }

This test is also based on changing the order $\nu^t(a)$  of
alphabet letters but the rule of the order change  differs from
(\ref{nu}). To describe the rule we first define $
\lambda^{t+1}(a)$ as a count of occurrences of $a$ in the word
$x_1\ldots x_{t-1}x_t .$ At each moment $t$ the alphabet letters
are ordered according to $\nu^t$ in such a way that, by
definition, for each pair of letters $a$ and $b$ $\nu^t(a) \prec
\nu^t(b)$ if $\lambda^t(a) \leq \lambda^t(b).$ For example, if $A=
\{a_1, a_2, a_3 \}$ and $x_1 x_2 x_3 = a_3 a_2 a_3$,  the possible
orders can be as follows: $\nu^1=(1, 2, 3),$ $ \nu^2=(3, 1, 2),$ $
\nu^3=(3, 2, 1),$ $ \nu^4=(3, 2,  1).$ In all other respects this
method coincides with the book stack. (The set of all indexes  $
\{1, \ldots, S \} $ is divided into $r  $ subsets,
 etc.)

Obviously,  after observing each letter $x_t$ the value
$\lambda^t(x_t)$ should be increased and the order $\nu^t$ should
be changed.  It is worth noting that there exist a data structure
and algorithm, which allow maintaining the alphabet letters
ordered in such a way that the number of operations spent is
constant, independently of the size of the alphabet. This data
structure was described by Moffat, 1999 and Ryabko, Rissanen,
2003.

\textbf{ 3.3. Two- faced processes and the choice of the block
length for a process testing }

There are quite many methods for testing $H_0$ and $H_1$, where
the bit stream is divided into words (blocks) of the length $s, s
\geq 1,$ and the sequence of the blocks $x_1x_2\ldots x_s$,
$x_{s+1}\ldots x_{2s},\ldots $ is considered as letters, where
each letter belongs to the alphabet $B_s = \{ 0,1 \}^s $ and has
the probability $2^{-s},$  if $H_0$ is true. For instance, both
above described tests, methods from Ryabko, Stognienko and Shokin
(2003) and quite  many other algorithms belong to this kind. That
is why the questions of choosing  the block length $s$ will be
considered here.

As it was mentioned in the introduction there exist  two-faced
processes, which, on the one hand, are far from being truly
random, but, on the other hand, they can be distinguished from
truly random only in the case when the block length $s$ is large.
From the information theoretical point of view the two- faced
processes can be simply described as follows. For a two- faced
process, which generates letters from  $ \{ 0,1 \}$, the limit
Shannon entropy is (much) less than 1 and, on the other hand, the
$s-$ order entropy ($h_s$) is maximal $(h_s =1$ bit per letter)
for relatively large $s.$

We describe two families of two- faced processes $T(k, \pi)$ and
$\bar{T}(k, \pi)$, where $k=1,2, \ldots,\,$ and $ \pi \in (0,1)$
are parameters. The processes $T(k,\pi)$ and $\bar{T}(k, \pi)$ are
Markov chains of the connectivity (memory) $k$, which generate
letters from $\{0,1\}$. It is convenient to define them
inductively. The process $T(1,\pi)$ is defined by conditional
probabilities $P_{T(1, \pi)}(0/0) = \pi, P_{T(1, \pi)}(0/1) =
1-\pi $ (obviously, $P_{T(1, \pi)}(1/0) =1- \pi, P_{T(1,
\pi)}(1/1) = \pi $). The process $\bar{T}(1,\pi)$ is defined by
$P_{\bar{T}(1, \pi)}(0/0) =1- \pi, P_{\bar{T}(1, \pi)}(0/1) = \pi
$. Assume that $T(k, \pi)$ and $\bar{T}(k, \pi)$ are defined and
describe $T(k+1, \pi)$ and $\bar{T}(k+1, \pi)$ as follows $$
P_{T(k+1, \pi)}(0/0u) = P_{T(k, \pi)}(0/u), P_{T(k+1, \pi)}(1/0u)
= P_{T(k, \pi)}(1/u), $$ $$ P_{T(k+1, \pi)}(0/1u) = P_{\bar{T}(k,
\pi)}(0/u), P_{T(k+1, \pi)}(1/1u) = P_{\bar{T}(k, \pi)}(1/u) ,$$
and, vice versa, $$ P_{\bar{T}(k+1, \pi)}(0/0u) = P_{\bar{T}(k,
\pi)}(0/u), P_{\bar{T}(k+1, \pi)}(1/0u) = P_{\bar{T}(k,
\pi)}(1/u), $$ $$ P_{\bar{T}(k+1, \pi)}(0/1u) = P_{T(k,
\pi)}(0/u), P_{\bar{T}(k+1, \pi)}(1/1u) = P_{T(k, \pi)}(1/u) $$
for each $u \in B_k$ (here $vu$ is a concatenation of the words
$v$ and $u$). For example, $$ P_{T(2,\pi)}(0/00) = \pi,
P_{T(2,\pi)}(0/01) = 1-\pi, P_{T(2,\pi)}(0/10) = 1-\pi,
P_{T(2,\pi)}(0/11) = \pi. $$ The following theorem shows that the
two-faced processes exist.

\textbf{Theorem 3.} { \it  For each $\pi \in (0,1) $ the s-order
Shannon entropy ($h_s$) of the processes $T(k, \pi)$ and
$\bar{T}(k, \pi)$ equals 1 bit per letter for $s=0,1,\ldots , k$
whereas the limit Shannon entropy ($h_\infty $) equals $ - (\pi
\log_2 \pi + (1- \pi) \log_2 (1-\pi) ).$ }

The proofs of the theorem is given in Appendix, but here we
consider examples of "typical" sequences of the processes
$T(1,\pi)$ and $\bar{T}(1,\pi)$ for $\pi$, say, 1/5. Examples are:
$ 010101101010100101...$ and $ 000011111000111111000.... .$ We can
see that each sequence contains approximately one half of 1's and
one half of 0's. (That is why the first order Shannon entropy is
1 per a letter.) On the other hand, both sequences do not look
like truly random, because they, obviously, have too long
subwords like either  $101010 ..$ or $000.. 11111.. .$ (In other
words, the second order Shannon entropy is much less than 1 per
letter.) Hence, if a randomness test is based on estimation of
frequencies of 0's and 1's only, then such a test will not be
able to find deviations from randomness.

So, if we revert to the question about the block length of tests
and take into account the existence of two- faced processes,  it
seems that the block length could be taken as large as possible.
But it is not so. The following informal consideration could be
useful for choosing the block length. The point is that
statistical tests can be applied if words from the sequence

\begin{equation}\label{s}
x_1x_2\ldots x_s, \:x_{s+1}\ldots x_{2s},\ldots,  \:x_{(m-1)s+1}
x_{(m-1)s+2}\ldots x_{m s} \end{equation}
 are repeated (at least a few times)
with high probability (here $m s $ is the sample length).
Otherwise, if all words in (\ref{s}) are unique (with high
probability) when $H_0$ is true, a sensible test cannot be
constructed basing on a division into $s-$letter words. So, the
word length $s$ should be chosen in such a way that some words
from the sequence (\ref{s}) are repeated with high probability,
when $H_0$ is true. So, now our problem can be formulated as
follows. There is a binary sequence $x_1x_2\ldots x_n$ generated
by the Bernoulli source with $P(x_i=0)= P(x_i=1)= 1/2 $ and we
want to find such a block length $s$ that the sequence (\ref{s})
with $ m= \lfloor n/s\rfloor, $ contains some repetitions (with
high probability). This problem is well known in the probability
theory and sometimes called as the birthday problem. Namely, the
standard statement of the problem is as follows. There are $S=
2^s$ cells and $m\, (=n/s)$ pellets. Each pellet is put in one of
the cells with the probability $1/S$. It is known in Probability
Theory that, if $m = c\, \sqrt{ S}, c >0$ then the average number
of cells with at least two pellets equals $c^2\, (1/2 + \circ
(1)\,),$ where $S$ goes to $\infty \,;$ see Kolchin, Sevast'yanov
and Chistyakov, 1976. In our case the number of cells with at
least two pellets is equal to the number of the words from the
sequence (\ref{s}) which are met two (or more) times. Having into
account that $S=2^s, m= n/s,$ we obtain from $m = c\, \sqrt{ S}, c
>0$ an informal rule for choosing the length of words
in (\ref{s}): \begin{equation}\label{sn} n \asymp s 2^{s/2}
\end{equation} where $n$ is the length of a sample $x_1x_2
... x_n,$ $s$ is the block length. If $s$ is much larger, the
sequence (\ref{s}) does not have repeated words (in case $H_0$ )
and it is difficult to build a a sensible test. On the other hand,
if $s$ is much smaller, large classes of the alternative
hypotheses cannot be tested (due to existence of the two-faced
processes). It is worth noting that it is impossible to have a
universal choice of $s,$ because it is impossible to avoid the
two- faced phenomenon. In other words this fact can be explained
basing on the following known result of Information Theory: it is
impossible to have guaranteed rate of code convergence
universally for all ergodic sources; see Bailey, 1976, Ryabko,
1984. That is why, it is impossible to choose a universal length
$s.$ On the other hand, there are many applications where the
word length $s$ can be chosen experimentally. (But, of course,
such experiments should be performed on the independent data.)

\section{The experiments }

In this part we describe some experiments carried out to compare
new tests with known ones. We will compare order test, book stack
test, tests which are based on standard data compression methods,
and tests from Rukhin and others, 2001. The point is that the
tests from Rukhin and others are selected basing on comprehensive
theoretical and experimental analysis and can be considered as the
state-of-the-art in randomness testing. Besides, we will also test
the method recently published  by Ryabko, Stognienko, Shokin,
(2004), because it was published later than the book of Rukhin
and others.

We used data generated by the  PRNG  "RANDU" (described in
Dudewicz and Ralley, 1981) and random bits from "The Marsaglia
Random Number CDROM", see: http://stat.fsu.edu/diehard/cdrom/ ).
RANDU is  a linear congruent generators (LCG), which is defined by
the following equality $$X_{n+1}=(A \: X_n+C)\: mod\, M \, ,$$
where $X_{n}$ is $n$-th generated number. RANDU is defined by
parameters $A=2^{16}+3  , C= 0 , M= 2^{31} , X_0 = 1.$ Those kinds
of sources of random data were chosen because random bits from
"The Marsaglia Random Number CDROM"  are considered as good random
numbers, whereas it is known that RANDU is not a good PRNG. It is
 known   that the  lowest digits of $X_n$ are "less random" than
the leading digits (Knuth, 1981), that is why in our experiments
with RANDU we extract an eight-bit word from each generated $X_i$
by formula $ \hat{X}_i = \lfloor X_i/2^{23} \rfloor .$

The behavior of the tests was investigated for files of different
lengths (see the tables below). We generated 100 different files
of each length and applied each  mentioned above test to each file
with level of significance 0.01 (or less, see below). So, if a
test is applied to a truly random bit sequence, on average 1 file
from 100 should be rejected. All results are given in the tables,
where integers in boxes are the number of rejected files (from
100). If a number of the rejections is not given for a certain
length and test, it means that the test cannot be applied for
files of such a length.

The table 1 contains information about testing of sequences of
different lengths generated by RANDU, whereas the table 2 contains
results of application of all tests to  5 000 000- bit sequences
either generated by RANDU or  taken from "The Marsaglia Random
Number CDROM".
 For example, the first number of the second row of
the table 1 is 56. It means that there were 100 files of the
length $5 \: 10^4$ bits generated by PRNG RANDU. When the Order
test was applied, the hypothesis $H_0$ was rejected 56 times from
100 (and, correspondingly, $H_0$ was accepted 44 times.) The first
number of the third line shows that $H_0$ was rejected 42 times,
when the Book stack test was applied to the same 100 files. The
third number of the second line shows that the hypothesis $H_0$
was rejected 100 times, when the Order test was applied for
testing of 100 $ 100000-$bit files generated by RANDU, etc.

Let us first give some comments about the tests, which are based
on popular data compression methods RAR and ARJ. In those cases we
applied each  method to a file and first estimated the length of
compressed data. Then we use the test
$\Gamma_{\alpha,\,\varphi}^{(n)}$ with the critical value
(\ref{cr}) as follows. The alphabet size $|A|= 2^8 = 256$, $ n
\log |A|$ is simply the length of file (in bits) before
compression, (whereas $n$ is the length in bytes). So, taking
$\alpha = 0.01,$ from (\ref{cr}), we see that  the hypothesis
about randomness ($H_0$) should be rejected, if the length of
compressed file is less than or equal to $ n \log |A| - 8$ bits.
(Strictly speaking, in this case $\alpha \leq 2^{-7} = 1/128.$)
So, taking into account that the length of computer files is
measured in bytes, this rule is very simple: if the $n-$byte file
is really compressed (i.e. the length of the encoded file is
$n-1$ bytes or less), this file is not random (and $H_0$ is
rejected). So, the following tables contain numbers of cases,
where files were really compressed.

Let us now give some comments about parameters of the considered
methods. As it was mentioned, we investigated all methods from the
book of Rukhin and others (2001), the test of Ryabko, Stognienko
and Shokin, 2004 (RSS test for short), the described above two
tests based on data compression algorithms,  the order tests and
the book stack test. For some tests there are parameters, which
should be specified. In such cases the values of parameters are
given in the table in the row, which  follows the test results.
There are some tests from the book of  Rukhin and others, where
parameters can be chosen from a certain  interval. In such cases
we repeated all calculations three times, taking the minimal
possible value of the parameter, the maximal one and the average
one. Then the data for the case when the number of rejections of
the hypothesis $H_0$ is maximal, is given in the table.

The choice of  parameters for RSS, the book stack test and the
order test was made on the basis of special experiments, which
were carried out for independent data. (Those algorithms are
implemented as a Java program and can be found on the internet,
see  $ http://web.ict.nsc.ru/\: \tilde{}\: rng/ $.) In all cases
such experiments have shown that for all three algorithms the
optimal blocklength is close to the one defined by informal
equality (\ref{sn}).

We can see from the tables that the new tests can detect
non-randomness more efficiently than the known ones. Seemingly,
the main reason is that RSS, book stack tests and order test deal
with such large blocklength as it is possible, whereas  many other
tests are focused on other goals. The second reason could be an
ability for adaptation. The point is that the new tests  can find
subwords, which are more frequent than others, and use them for
testing, whereas many other tests are looking for particular
deviations from randomness.

In conclusion, we can say that the  obtained results show that the
new tests, as well as the ideas of Information Theory in general,
can be useful tools for randomness testing.

\begin{table}[h]
\caption{ Number of files generated by PRNG RANDU and recognized
as non-random for different tests and different file lengths (in
bits). }
\begin{center}
 \begin{tabular}{|c|c|c|c|c|}

\hline \rule{0pt}{2.8ex}Name of test/Length of file
  &$5 \: 10^4$ &$10^5$&$ 5 \:10^5$& $10^6$\\
\hline \rule{0pt}{2.3ex}Order test &56&100&100&100\\
 \rule{0pt}{2.3ex}Book stack &42&100&100&100\\
\cline{2-5}
 \rule{0pt}{2.3ex}{\it parameters for both tests} &\multicolumn{4}{|c|} { s=20, $|A_1|=5\sqrt{2^{s}}$}\\
\hline
 \rule{0pt}{2.3ex}RSS &4&75&100&100\\
\cline{4-5}
 \rule{0pt}{2.3ex}{\it parameters} &s=16&s=17&\multicolumn{2}{|c|} {s=20}\\
\hline
\rule{0pt}{2.3ex} RAR &0&0&100&100\\
\rule{0pt}{2.3ex} ARJ &0&0&99&100\\
\hline \rule{0pt}{2.3ex}Frequency& 2&1&1&2\\ \hline
\rule{0pt}{2.3ex}Block Frequency &1&2&1&1\\
 \rule{0pt}{2.3ex}{\it parameters} &M=1000&M=2000&$M=10^5$&M=20000\\
\hline \rule{0pt}{2.3ex}Cumulative Sums&2&1&2&1\\ \hline
 \rule{0pt}{2.3ex}Runs&0&2&1&1\\
\hline
  \rule{0pt}{2.3ex}Longest Run of Ones &0&1&0&0\\
\hline
  \rule{0pt}{2.3ex}Rank &0&1&1&0\\
\hline \rule{0pt}{2.3ex}Discrete Fourier Transform &0&0&0&1\\
\hline
\rule{0pt}{2.3ex}NonOverlapping Templates &--&--&--&2\\
 \rule{0pt}{2.3ex}{\it parameters} &&&&m=10\\
\hline\rule{0pt}{2.3ex} Overlapping Templates&--&--&--&2\\
 \rule{0pt}{2.3ex}{\it parameters} &&&&m=10\\
\hline
\rule{0pt}{2.3ex}Universal Statistical &--&--&1&1\\
 \rule{0pt}{2.3ex}{\it parameters} &&&L=6&L=7\\
 \rule{0pt}{2.3ex} &&& Q=640&Q=1280\\
\hline
\rule{0pt}{2.3ex}Approximate Entropy&1&2&2&7\\
 \rule{0pt}{2.3ex}{\it parameters} &m=5&m=11&m=13&m=14\\
\hline \rule{0pt}{2.3ex}Random Excursions &--&--&--&2\\ \hline
\rule{0pt}{2.3ex}Random Excursions Variant&--&--&--&2\\ \hline
\rule{0pt}{2.3ex}Serial &0&1&2&2\\
 \rule{0pt}{2.3ex}{\it parameters} &m=6&m=14&m=16&m=8\\
\hline \rule{0pt}{2.3ex}Lempel-Ziv Complexity&--&--&--&1\\ \hline
\rule{0pt}{2.3ex}Linear Complexity &--&--&--&3\\
 \rule{0pt}{2.3ex}{\it parameters} &&&&M=2500\\
\hline
 \end{tabular}
 \end{center}
 \end{table}

\begin{table}[!hbt]
  \caption{ Number of $5 \,000 \,000-$ bit  files
generated by PRNG RANDU  and random, which are recognized as
non-random. }
\begin{center}
 \begin{tabular}{|c|c|c|}
\hline
 \rule{0pt}{2.8ex}Name of test/ Kind of file
  &$ \: RANDU$ &$random $\\

\hline \rule{0pt}{2.3ex}Order test &100&3\\
 \rule{0pt}{2.3ex}Book stack &100&0\\
\cline{2-3}

 \rule{0pt}{2.3ex}{\it parameters for both tests} &\multicolumn{2}{|c|} {s=24, $|A_1|=5\sqrt{2^{s}}$}\\
\hline
 \rule{0pt}{2.3ex}RSS &100&1\\
\cline{2-3}

 \rule{0pt}{2.3ex}{\it parameters} &s=24&s=24\\
\hline
\rule{0pt}{2.3ex} RAR &100&0\\
\rule{0pt}{2.3ex} ARJ &100&0\\

\hline \rule{0pt}{2.3ex}Frequency& 2&1\\
\hline \rule{0pt}{2.3ex}Block Frequency &2&1\\
 \rule{0pt}{2.3ex}{\it parameters} &$M=10^6$&$M=10^5$\\
\hline

\rule{0pt}{2.3ex}Cumulative Sums&3&2\\
\hline

 \rule{0pt}{2.3ex}Runs&2&2\\
\hline

  \rule{0pt}{2.3ex}Longest Run of Ones &2&0\\
\hline

  \rule{0pt}{2.3ex}Rank &1&1\\
\hline

\rule{0pt}{2.3ex}Discrete Fourier Transform &89&9\\
\hline

\rule{0pt}{2.3ex} NonOverlapping Templates&5&5\\
 \rule{0pt}{2.3ex}{\it parameters}&m=10&m=10\\

\hline

\rule{0pt}{2.3ex} Overlapping Templates&4&1\\
 \rule{0pt}{2.3ex}{\it parameters} &m=10&m=10\\

\hline

\rule{0pt}{2.3ex}Universal Statistical &1&2\\
 \rule{0pt}{2.3ex}{\it parameters} &L=9&L=9\\
 \rule{0pt}{2.3ex} &Q=5120&Q=5120\\

\hline

\rule{0pt}{2.3ex}Approximate Entropy&100&89\\
 \rule{0pt}{2.3ex}{\it parameters} &m=17&m=17\\

\hline

\rule{0pt}{2.3ex}Random Excursions &4&3\\
\hline

\rule{0pt}{2.3ex}Random Excursions Variant&3&3\\
\hline

\rule{0pt}{2.3ex}Serial &100&2\\
 \rule{0pt}{2.3ex}{\it parameters} &m=19&m=19\\

\hline

\rule{0pt}{2.3ex}Lempel-Ziv Complexity&0&0\\
\hline

\rule{0pt}{2.3ex}Linear Complexity &4&3\\
 \rule{0pt}{2.3ex}{\it parameters} &M=5000 & M=2500 \\

\hline
 \end{tabular}
 \end{center}
 \end{table}

\section{Appendix. }

\emph{Proof} of Theorem 1. First we estimate the number of words
$\varphi_n(u)  $  whose length is less than or equal to an integer
$\tau$. Obviously, at most one word can be encoded by the empty
codeword, at most two words by the words of the length 1, ..., at
most $2^i$ can be encoded by the words of length $i,$ etc. Having
taken into account that the codewords $\varphi_n(u) \neq
\varphi_n(v)$ for different $u$ and $v$, we obtain the inequality
$$ | \{ u: |\varphi_n(u) | \leq \tau \} | \leq \sum_{i=0}^\tau 2^i
= 2^{\tau+1}- 1. $$ From this inequality and (\ref{cr}) we can see
that the number of words from the set $ \{A^n \} ,$ whose
codelength is less than or equal to $t_\alpha = n \log |A| - \log
(1/ \alpha) - 1 ,$ is not greater than $ 2^{n \log |A| - \log (1/
\alpha)}.$ So, we obtained that $$ | \{ u: |\varphi_n(u) | \leq
t_\alpha \} | \leq \alpha |A|^n .$$ Taking into account that all
words from $A^n$ have equal probabilities if $H_0$ is true, we
obtain from the last inequality, (\ref{cr}) and the description of
the test $\Gamma_{\alpha,\varphi}^{(n)}$ that $$ Pr \{
|\varphi_n(u) | \leq t_\alpha | \} \leq  (\alpha |A|^n / |A|^n ) =
\alpha $$ if $H_0$ is true. The theorem is proved.

\emph{Proof} of Claim 1. The proof is based on a direct
calculation of the probability of  rejection for a case  where
$H_0$ is true. From the description of the test
$\Upsilon_{\alpha,\varphi}^{(n)}$ and definition of $g$ (see
(\ref{s})) we obtain the following chain of equalities.

$$ Pr \{ H_0\: is \:rejected \,\}= Pr \{\, |\varphi_n(u) | \leq g \}
$$ $$+\, Pr \{ |\varphi_n(u) | = g+1 \}\: (\:
 1\:-\, (\sum_{j=1}^{g+1} |A_j|\, - \, \alpha
|A|^n\,) / |A_{g+1}|\,)\,) $$ $$= \frac{1}{A^n} \: ( \sum_{j=0}^g
|A_j|\:+ \: |A_{g+1}|\: (\:
 1\:-\, (\sum_{j=1}^{g+1} |A_j|\, - \, \alpha
|A|^n\,) / |A_{g+1}|\,)\,)= \alpha .$$ The claim is proved.

\emph{Proof} of Claim 2. We can think that $\hat{t}_\alpha$ in
(\ref{cr2}) is an integer. (Otherwise, we obtain the same test
taking $\lfloor\hat{t}_\alpha\rfloor$ as a new critical value of
the test.) From the  Kraft inequality (\ref{KRAFT}) we obtain that
$$ 1\geq \sum_{u \in A^n } 2^{- |\varphi_n (u)|} \geq | \{u: |\,
\varphi_n (u)|\leq \hat{t}_\alpha  \} | \: \:2^{-\hat{t}_\alpha}.
$$ This inequality and (\ref{cr2}) yield: $$ | \{u: |\, \varphi_n
(u)|\leq \hat{t}_\alpha  \} | \leq \alpha |A|^n. $$ If $H_0$ is
true then the probability of each $u \in A^n $ equals $|A|^{-n} $
and from the last inequality we obtain that $$ Pr \{ |\varphi (u)
| \leq \hat{t}_\alpha  \} = |A|^{-n} \: | \{u: |\, \varphi_n
(u)|\leq \hat{t}_\alpha  \} | \leq \alpha , $$ if $H_0$ is true.
The claim is proved.

\emph{Proof} of Theorem 3. We prove the theorem for the process
$T(k, \pi),$  but this proof is valid for $\bar{T}(k, \pi),$ too.
 First we show that
\begin{equation}\label{a3} p^*(x_1...x_d)=2^{-d}, \end{equation}
 $ (x_1...x_{d}) \in \{ 0,1 \}^{d}, $ $d =1, ... , k,$ is a
stationary distribution for the processes $T(k, \pi)$ (and
$\bar{T}(k, \pi)$) for all $k=1,2, \ldots $ and $ \pi \in (0,1)$.
 For any values of $k, k \geq 1,$ (\ref{a3}) will be proved if we
show that the system of equations $$ P_{T(k, \pi)}(x_1...x_d)=
P_{T(k, \pi)}(0x_1...x_{d-1})\, P_{T(k,
\pi)}(x_d/0x_1...x_{d-1})\: $$ $$ +\,P_{T(k,
\pi)}(1x_1...x_{d-1})\, P_{T(k, \pi)}(x_d/1x_1...x_{d-1}) $$ has
the solution $p(x_1...x_d)=2^{-d}$, $ (x_1...x_{d}) \in \{ 0,1
\}^{d}, $ $d =1,2, \ldots, k $. It can be easily seen for $d =
k,$ if we take into account that, by definition of $T(k, \pi)$
and $\bar{T}(k, \pi)$, the equality $P_{T(k,
\pi)}(x_k/0x_1...x_{k-1})\: +\, P_{T(k,
\pi)}(x_k/1x_1...x_{k-1})=1 $ is valid for all $ (x_1...x_{k})
\in \{ 0,1 \}^{k} $. From this equality and the law of total
probability we immediately obtain (\ref{a3}) for $d < k.$

Let us prove the second claim of the theorem. From the definition
$T(k, \pi)$ and $\bar{T}(k, \pi)$ we can see that either $P_{T(k,
\pi)}(0/x_1...x_{k})= \pi,\, P_{T(k, \pi)}(1/x_1...x_{k})=1-\pi$
or $P_{T(k, \pi)}(0/x_1...x_{k})=1- \pi,\, P_{T(k,
\pi)}(1/x_1...x_{k})=\,\pi$. That is why $h(x_{k+1}/x_1...x_{k}) =
- (\pi \log_2 \pi + (1- \pi) \log_2 (1-\pi) )$ and, hence,
$h_\infty = - (\pi \log_2 \pi + (1- \pi) \log_2 (1-\pi) )$. The
theorem is proved.

\section{Acknowledgment } The authors wish to thank  one of anonymous
reviewers for information about a unpublished thesis of David
Harold Bailey.


\end{document}